\documentclass[aps,prb,twocolumn,showpacs,floatfix]{revtex4-1}  

\usepackage[dvips]{graphicx}
\usepackage{color}
\usepackage{amsmath}
\usepackage{ifsym}
\usepackage{amssymb}

\newcommand{\degC}{$^{\circ}$C}
\newcommand{\degrees}{$^{\circ}$}
\newcommand{\FeCoSi}{Fe$_{1-x}$Co$_{x}$Si}

\begin{document}
\title{Strain-induced effects on the magnetic and electronic properties of epitaxial \FeCoSi\ thin films}

\author{P.~Sinha}\email[email:~]{pyps@leeds.ac.uk}
\affiliation{School of Physics \&\ Astronomy, University of Leeds, Leeds, LS2 9JT, UK}

\author{N.~A.~Porter}
\affiliation{School of Physics \&\ Astronomy, University of Leeds, Leeds, LS2 9JT, UK}

\author{C.~H.~Marrows}\email[email:~]{c.h.marrows@leeds.ac.uk}
\affiliation{School of Physics \&\ Astronomy, University of Leeds, Leeds, LS2 9JT, UK}


\begin{abstract}
  We have investigated the Co-doping dependence of the structural, transport, and magnetic properties of $\epsilon$-\FeCoSi\ epilayers grown by molecular beam epitaxy on silicon (111) substrates. Low energy electron diffraction, atomic force microscopy, X-ray diffraction, and high resolution transmission electron microscopy studies have confirmed the growth of phase-pure, defect-free $\epsilon$-\FeCoSi\ epitaxial films with a surface roughness of $\sim 1$~nm. These epilayers are strained due to lattice mismatch with the substrate, deforming the cubic B20 lattice so that it becomes rhombohedral. The temperature dependence of the resistivity changes as the Co concentration is increased, being semiconducting-like for low $x$ and metallic-like for $x \gtrsim 0.3$. The films exhibit the positive linear magnetoresistance that is characteristic of $\epsilon$-\FeCoSi\ below their magnetic ordering temperatures $T_\mathrm{ord}$, as well as the huge anomalous Hall effect of order several $\mu \Omega$cm. The ordering temperatures are higher than those observed in bulk, up to 77~K for $x = 0.4$. The saturation magnetic moment of the films varies as a function of Co doping, with a contribution of $\sim 1~\mu_{\rm B}$/ Co atom for $x \lesssim 0.25 $. When taken in combination with the carrier density derived from the ordinary Hall effect, this signifies a highly spin-polarised electron gas in the low $x$, semiconducting regime.
\end{abstract}

\date{\today}
\maketitle

\section{introduction}

The rich behaviour shown by ferromagnetic semiconductors arise from an interesting interplay of their electronic density of states and magnetic interactions within the crystal structure, offering new possibilities for spintronics.\cite{Ohno2010} Whilst most magnetic semiconductors to date are based on compound or oxide materials, the transition metal monosilicides are promising candidates in that they are based on silicon, by far the most common commercial semiconductor. These materials crystallize in cubic B20 structure, the $\epsilon$-phase, and which belongs to the space group $P2_1 3$.\cite{AlSharif2001} They are continuously miscible with each other and form an isostructural series compounds with endmembers MnSi (a metallic helimagnet), FeSi (a paramagnetic narrow-gap semiconductor), and CoSi (a metallic diamagnet).\cite{Manyala2000} They have been studied for many years as they exhibit wide variety of different aspects of condensed matter physics including paramagnetic anomalies,\cite{Wertheim1965,Jaccarino1967} strongly correlated/Kondo insulator-like behaviour,\cite{Schlesinger1993,Paschen1997,DiTusa1997,Aeppli1999} non-Fermi liquid behaviour,\cite{Pfleiderer2001,Manyala2008,ritz_formation_2013} unusual magnetoresistance,\cite{Manyala2000,Onose2005,Porter2012} and helical magnetism\cite{Beille1981,Beille1983,Uchida2006,Grigoriev2009} with skyrmion phases\cite{Muhlbauer2009,Munzer2010,Yu2010,milde_unwinding_2013} that have associated topological Hall effects.\cite{Lee2009,Neubauer2009,ritz_giant_2013}

Almost all work to date on the monosilicide materials has been carried out using bulk single crystal samples. For technological applications, thin films that can be patterned into devices with conventional planar processing techniques are required. Epilayers of the helimagnetic metal MnSi have been grown by using molecular beam epitaxy (MBE) by Karhu \textit{et al.},\cite{Karhu2010,Karhu2011,Karhu2012} Li \textit{et al.},\cite{li_robust_2013} and Engelke \textit{et al.}\cite{Engelke2012} The properties are broadly comparable to those of the bulk material, including the presence of chiral magnetism\cite{Karhu2011} and a topological Hall effect.\cite{li_robust_2013} Other monosilicides have received less attention to date. The family of alloys \FeCoSi\ should be of particular interest for spintronics: whilst both endmembers are non-magnetic, magnetic ordering is evident at almost all intermediate values of $x$.\cite{Manyala2000} For low doping levels of Co in the semiconducting parent FeSi, a magnetic semiconductor with a half-metallic state is expected.\cite{Manyala2000,Guevara2004} Polycrystalline thin films of \FeCoSi\ have been grown by pulsed laser deposition,\cite{Manyala2009} and sputtering,\cite{Morley2011} but with properties that fall short of those in single crystal samples due to microstructural disorder and lack of phase purity.

Here we report on the properties of epitaxial $\epsilon$-\FeCoSi\ layers grown on commercial (111) Si substrates, across the doping range $0 \leq x \leq 0.5$, using the growth methods we have previously developed.\cite{Porter2012} The films are phase pure, with a B20 lattice that is distorted by biaxial in-plane epitaxial strain to have a rhombohedral unit cell. Although \FeCoSi\ is known to possess a helimagnetic ground state,\cite{Beille1981,Beille1983,Uchida2006,Grigoriev2009} we focus here on the properties in fields large enough to generate a uniformly magnetized ferromagnetic state, which are modest in size. We find that these epilayers display the full range of properties expected of this material, including a characteristic temperature dependence of resistivity,\cite{Onose2005}, positive linear magnetoresistance,\cite{Manyala2000,Onose2005}, and a very large anomalous Hall effect.\cite{Manyala2004} Measurements of the number of Bohr magnetons ($\mu_\mathrm{B}$) of magnetic moment and electron-like carriers per Co indicate the presence of a highly spin-polarised electron gas in the low doping ($x \lesssim 0.25$) regime,\cite{Manyala2000,Morley2011} where the half-metallic state is expected.\cite{Guevara2004} Nevertheless, the presence of epitaxial strain, giving rise to an expanded unit cell volume, leads to some quantitative changes, the most prominent of which is a substantial enhancement of the magnetic ordering temperature with respect to bulk crystals. These epilayers are suitable for patterning into nanostructures that may find use as spin injectors into silicon\cite{Min2006,Appelbaum2007,Huang2007} or exploit the chiral nature of the magnetism at low fields in skyrmion-based devices.\cite{Kiselev2011,Fert2013,Lin2013}

\section{Growth and structural characterisation}

\begin{figure}
  \includegraphics[width=8cm]{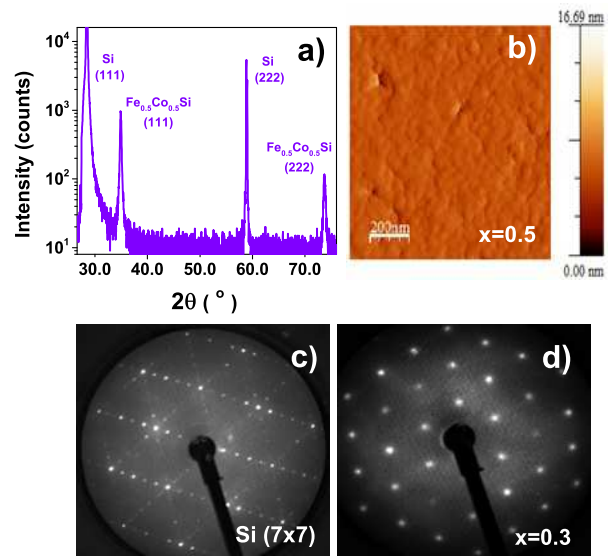}
  \caption{(Color online) Structural characterisation of the 50 nm thick \FeCoSi\ epilayers. (a) XRD spectrum of a $x = 0.5$ film, illustrating the phase purity of the B20 structure and the (111) epitaxial orientation of the film. (b) Atomic force micrograph of the top surface of an \FeCoSi\ epilayer with  $x = 0.5$. (c) LEED pattern of an annealed Si (111) substrate prior to film growth. The $7 \times 7$ surface reconstruction is evident. (d) LEED pattern from an  \FeCoSi\ film $x = 0.3$, demonstrating epitaxial growth in the (111) orientation. \label{fig:TMS_structure}}
\end{figure}

The \FeCoSi\ thin films were prepared by simultaneous co-evaporation of Fe, Co, and Si by MBE on a lightly n-doped silicon (111) substrates with $2000$-$3000~\Omega$cm resistivity at room temperature. The level of Co-doping $x$ of the various \FeCoSi\ films was determined by controlling the individual rates of incoming flux. We adopted the growth protocol described by Porter \textit{et al.} in Ref. \onlinecite{Porter2012}. The base pressure of the growth chamber remained within the range $2.8$-$4.8\times 10^{-11}$~mbar. Prior to the deposition of the film, the substrates were annealed at 1200\degC\ until a well ordered $7 \times 7$ reconstructed Si (111) surface was obtained. A low energy electron diffraction pattern demonstrating this reconstruction is shown in Fig.~\ref{fig:TMS_structure}(c). The films were then grown by depositing a seed layer of Fe of $\sim 5.4$~\AA\ thickness at room temperature, followed by the deposition of a $\sim 50$~nm thick \FeCoSi\ layer at a net flux rate of $\sim 0.4$~\AA/s at 400~\degC\ . The films were then further annealed at 400\degC\ for 15 minutes, before being allowed to cool to room temperature for further characterisation.

The films grow in the (111) orientation and are $\epsilon$-phase pure, as can be seen from the Cu $K_\alpha$ X-ray diffraction (XRD) spectrum shown in Fig.~\ref{fig:TMS_structure}(a). In-plane epitaxy of the \FeCoSi\ films is seen to be achieved by a 30\degrees\ in-plane rotation of the surface unit cell with respect to the Si, such that the \FeCoSi\ [11\={2}] direction is aligned parallel to Si [1\={1}0], demonstrated by the LEED pattern of a completed epilayer in Fig.~\ref{fig:TMS_structure}(d). Atomic force microscopy (AFM) was used to map the surface topography of the films: an representative micrograph is shown in Fig.~\ref{fig:TMS_structure}(b). The root mean square (rms) roughness of the films were estimated from these images to be around 1~nm.

\begin{figure}
  \includegraphics[width=5cm]{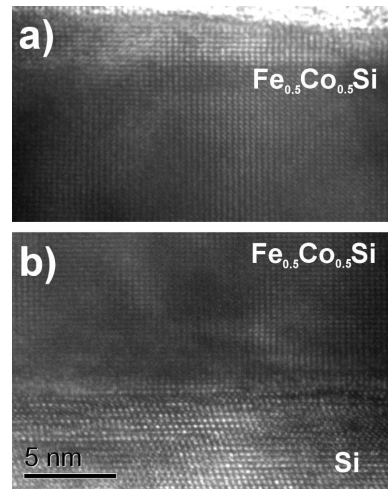}
  \caption{HRTEM of an \FeCoSi\ epilayer with $x = 0.5$ on the [112] zone axis, showing the upper (a) and lower (b) interfaces. \label{fig:TMS_TEM}}
\end{figure}

For further structural verification, high resolution transmission electron microscopy (HRTEM) and energy dispersive X-ray analysis (EDX) were carried out on cross-section specimens prepared by focussed ion beam (FIB). Fig.~\ref{fig:TMS_TEM}(a) and (b) show the top and bottom interfaces of a \FeCoSi\ film with $x = 0.5$. The films look well-ordered throughout and epitaxial growth can be observed with the orientation (111)\FeCoSi $\|$(111)Si~:~[11\={2}]\FeCoSi $\|$[1\={1}0]Si. Sample cross sections were mapped with EDX which confirmed the homogeneous distribution and chemical composition of the films. In-plane (110) lattice parameters were determined from the HRTEM images, which we discuss below.

\section{Strain characterisation}

Heteroepitaxy gives rise to strained growth of films as a result of the lattice mismatch between substrate and the film. The lattice parameter of Si is 5.431~\AA, whilst that of  bulk FeSi is 4.482~\AA. It is to accommodate this large difference that the film grows with the 30\degrees\  in-plane rotation demonstrated above by LEED (see Fig.~\ref{fig:TMS_structure}(c) and ~\ref{fig:TMS_structure}(d)) and HRTEM (Fig.~\ref{fig:TMS_TEM}(a) and ~\ref{fig:TMS_TEM}(b)). This gives rise to an in-plane lattice mismatch of $5.6\%$ at the interface. Inspection of the LEED patterns shows that this is relaxed to $\sim 3.7\%$ at the surface of a 50 nm thick film (see above). The heteroepitaxy induces biaxial tensile strain in the in-plane directions of the \FeCoSi\ layers, with corresponding compression in the out-of-plane direction, which distorts the cubic B20 lattice to have a rhombohedral form.

\begin{figure}
  \includegraphics[width=8cm]{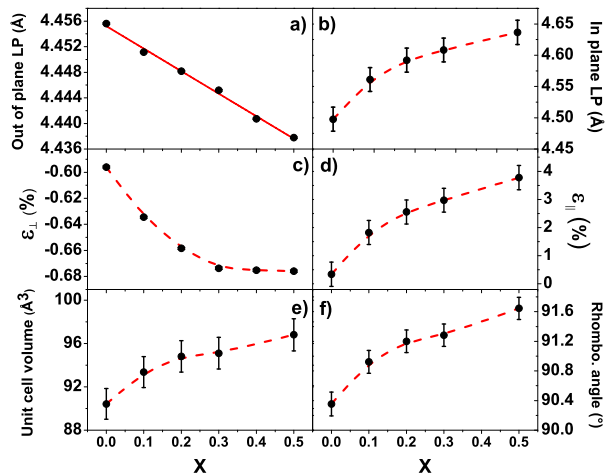}
  \caption{(Color online) Strain analysis. (a) Out-of-plane lattice parameter (LP) $a^{111}$ of \FeCoSi\ films based on data from XRD. (b) In-plane lattice parameter (LP) $a^{110}$ at the surface of the film, based on data from LEED. (c) Out-of-plane of strain in the unit cell. (d) In-plane strain in the unit cell. (e) Rhombohedral unit cell volume as a function of $x$. (f) Rhombohedral angle as a function of $x$. The solid lines are linear best fits, the dashed lines are guides to the eye. \label{fig:strain_2p}}
\end{figure}

The position of the \FeCoSi\ [111] and [222] Bragg peaks, obtained from $\theta$-$2\theta$ high angle XRD scans, were used to determine the out-of-plane [111] lattice parameter of \FeCoSi\ films using the Bragg law. In order to make quantitative comparisons of our samples, we define the parameter $a^{hkl}$, the lattice constant, assuming a cubic unit cell, that is determined from a measured interplanar spacing $d^{hkl}$ associated with a particular set of lattice planes $(hkl)$. A systematic decrease in out-of-plane lattice constant, $a^{111}$ is observed with increasing Co content $x$ in the films, as shown in Fig.~\ref{fig:strain_2p}(a). The linear variation of the out-of-plane lattice parameter with $x$ shows that Vegard's law is followed, as is the case in bulk crystals of this material.\cite{Shinoda:1972} However, there is also the large in-plane lattice mismatch with the Si substrate that was discussed above in the case of thin films. The in-plane lattice parameter $a^{110}$ at the surface of the \FeCoSi\ films, shown in Fig.~\ref{fig:strain_2p}(b), varies from $4.45\pm ~0.02$~\AA\ for $x=0$ to $4.64\pm 0.02 $~\AA\ for $x=0.5$, as determined from analysis of the LEED patterns, using the ($7 \times 7$) reconstructed Si (111) pattern to provide a calibration. Overall we see that the in-plane lattice parameter of epitaxial \FeCoSi\ is larger than the corresponding out-of-plane lattice parameter and is closer to that of Si (5.431\AA). The variation with $x$ is plotted in Fig.~\ref{fig:strain_2p}(b).

Based on data from Fig.~\ref{fig:strain_2p}(a)~and~\ref{fig:strain_2p}(b), the out-of-plane compressive, $\varepsilon_\perp$, and in-plane tensile, $\varepsilon_\|$, strains in the crystal structure were calculated, with the results shown in Fig.~\ref{fig:strain_2p}(c) and (d), using the following expression:
\begin{equation}
\varepsilon^{hkl} = \frac{a^{hkl}_\mathrm{epi}-a^{hkl}_\mathrm{bulk}}{a^{hkl}_\mathrm{bulk}},
\end{equation}
where $a^{hkl}_\mathrm{epi} $ is the lattice parameter as measured for a given epilayer and $a^{hkl}_\mathrm{bulk}$ is the corresponding lattice parameter in the bulk \citep{Manyala2004}. In both the cases strain follows a nonlinear relationship with the Co-doping level $x$. For higher values of $x$ the out-of-plane lattice constant is more compressed, whilst the in-plane lattice is extended.

\begin{figure}
  \includegraphics[width=8cm]{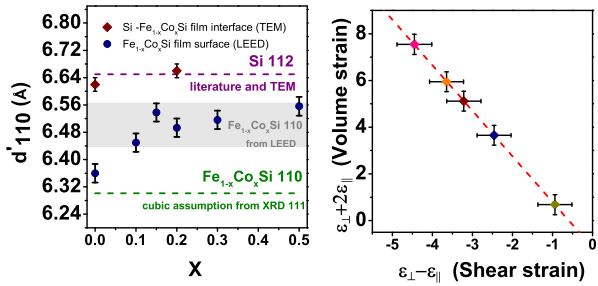}
  \caption{(Color online) Epitaxial strain analysis. (a) Comparison of evolution of unit cell face diagonal $d^\prime_{110}$ of \FeCoSi\ films  as a function of cobalt content from data obtained by LEED, TEM and theoretical prediction. b) Variation of volume strain with shear strain for various Co doping in \FeCoSi\ films. The dashed line is a straight line best fit to the data.  \label{fig:strain_3p}}
\end{figure}

The different methods we have used to determine the lattice constants give information about different parts of the film. Using the TEM images as shown in Fig.~\ref{fig:TMS_TEM}(b) it is possible to determine the lattice constant of the \FeCoSi\ near the Si substrate. In Fig.~\ref{fig:strain_3p}(a), we plot the unit cell face diagonal $d^\prime_{110}$ for selected values of $x$ as obtained from TEM. For $x=0$ and $x=0.2$, $d^\prime_{110}$ is measured to be  $6.62\pm ~0.02$~\AA\ and $6.66\pm ~0.02$~\AA\ respectively. These values are seen to match well to the Si (112) face diagonal (6.6501~\AA), which it must for heteroepitaxial growth. Our LEED data are surface sensitive, however. Measuring $d^\prime_{110}$ from our LEED patterns shows considerable variation with $x$ (Fig.~\ref{fig:strain_3p}(a)). For $x=0$, there is a good match to the bulk value for this crystallographic distance, if we assume a cubic crystal structure. We can conclude from this comparison that the \FeCoSi\ films are strained at the Si interface to adapt to the lattice constant of Si substrate. At greater distances from the interface with the substrate, the lattice relaxes throughout the 50~nm film thickness, and adapts to its own strained lattice constant for a rhombohedral crystal structure which is somewhere in between that of Si and the \FeCoSi\ cubic assumption of crystal structure. The variation of volume strain with shear strain in \FeCoSi\ film is shown in the Fig.~\ref{fig:strain_3p}(b) for various Co doping ranging from $x=0$ to $x=0.5$. The linearity in the relationship confirms that the epitaxial strain in \FeCoSi\ film changes only the angle of the unit cell as shown in Fig.~\ref{fig:strain_2p}(f) and that there are no structural phase changes associated with the strain. Thus, even though the strained \FeCoSi\ films have rhombohedral unit cell but they are phase pure as shown in the Fig.~\ref{fig:TMS_structure}(a).

Knowledge of the in-plane and out-of-plane lattice constants give a full determination of the geometry of the rhombohedral unit cell. The volume of the unit cell as function of $x$ is plotted in Fig.~\ref{fig:strain_2p}(e). The unit cell volume increases in a monotonic but non-linear fashion with $x$. We have also calculated the variation of the rhombohedral angle as a function the varying Co doping, shown in Fig.~\ref{fig:strain_2p}(f). The angle increases from little more than $90$\degrees\ for $x=0$  to $\sim 92$\degrees\ for $x=0.5$. Since the in-plane strain is determined from LEED, these values apply close to the top surface of the epilayer. These changes in unit cell geometry induced by  epitaxial strain can be expected to give rise to modifications to various properties such as the band structure, density of states, transport properties, magnetization and magnetic anisotropy, which we will explore in remainder of the paper.

\section{Transport Properties}\label{sec:rho}

The transport properties of our \FeCoSi\ films were measured in a gas-flow cryostat with a base temperature of 1.4~K capable of applying magnetic fields of up to 8~T. The films were patterned into Hall bars which were $5~\mu$m wide using optical lithography, etched by Ar ion milling, and bonded onto a chip carrier for measurement.

\begin{figure}
  \includegraphics[width=8.5cm]{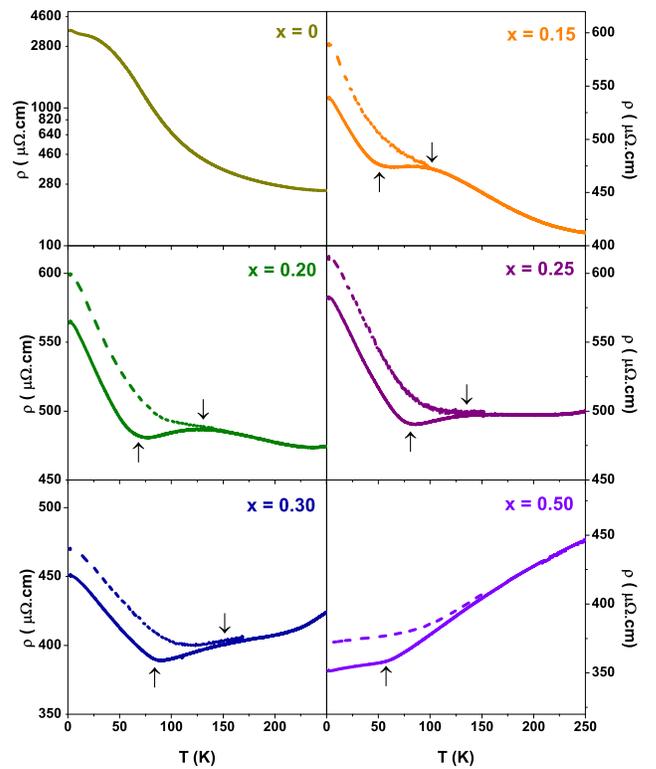}
  \caption{(Color online) Temperature dependence of resistivity in $\sim$ 50 nm films of \FeCoSi\ in magnetic fields of 0~T (solid lines) and 8 T (dashed lines). Increasing cobalt concentration $x$ changes the temperature coefficient of resistivity from negative (semiconductor-like) for $x = 0$ to positive (metallic-like) for $x = 0.5$, with mixed behavior seen for intermediate values of $x$. $\uparrow$ and $\downarrow$ illustrate respectively temperatures of minima, $T_{res}$, and maxima in the resistivity.\label{fig:TMS_RvT}}
\end{figure}

Measurements of the electrical resistivity $\rho(T,H)$ of the films as a function of temperature $T$ and magnetic field $H$ applied perpendicular to the sample plane are shown in Fig.~\ref{fig:TMS_RvT}. A bias current of $30~\mu$A was used. The solid lines show the $\rho(T)$ in absence of magnetic field and the dashed lines show $\rho(T)$ in presence of an 8~T magnetic field. Fig.~\ref{fig:TMS_RvT}(a) shows the resistivity variation of an FeSi film. FeSi is a narrow band-gap semiconductor,\cite{Jaccarino1967} and upon decreasing the temperature the resistivity increases reaching $3700~\mu\Omega$cm at 1.4K. We determined the band-gap of the epitaxial FeSi to be $\Delta = 30.1 \pm 0.2$~meV using the following relation:
\begin{equation}
\ln \rho \propto \left( \frac{\Delta}{2k_\mathrm{B} T} \right),
\end{equation}
where $k_B$ is the Boltzmann constant, fitted to the high temperature data (above about 50~K).

Doping FeSi with Co introduces electron-like carriers and a lowered resistivity. At the opposite extreme, the $\rho(T)$ relation for the film with $x = 0.5$ has a metallic form, shown in Fig.~\ref{fig:TMS_RvT}(f), increasing with $T$ for all temperatures. Intermediate values of $x$ yield hybrid $\rho(T,0)$ dependences, with a gradual crossover from semiconductor-like to metal-like behavior as $x$ rises. For these values of $x$ the $\rho(T,0)$ curve is often non-monotonic, combining regions with both positive and negative temperature coefficients of resistance. The curves are similar to those measured for bulk crystals at a qualitative level,\cite{Manyala2000,Onose2005} but differ quantitatively.

In the intermediate doping regime ($0.15<x<0.3$), we observe some distinctive features such as points of local maximum ($T_\mathrm{max}$) and minimum ($T_\mathrm{res}$) resistivity that vary with the degree of Co doping. For instance, in Fig.~\ref{fig:TMS_RvT}(b) (for $x=0.15$) we observe a broad maximum in $\rho$ around 125~K. As the Co doping increases this maximum shifts towards higher temperatures, reaching 175~K for $x=0.3$, then becoming less pronounced until it vanishes for $x=0.5$. The observed broad maximum is a feature reminiscent of the narrow band-gap semiconducting parent compound FeSi \cite{Onose2005}. The maxima and associated temperature shift can be explained in the framework of epitaxial strain and Co doping. Substituting Co for Fe not only introduces volume strain (as previously shown in Fig.~\ref{fig:strain_3p}(b)), but also changes the band structure, resulting in a broadening of bands and reduced band gap.\cite{Forthaus} Thus, increased Co doping provides more carriers to be available for conduction, giving rise to the hybrid semiconducting-metallic behaviour that we see. It is the competition between the temperature dependence of mobility, importance of thermally activated carriers (particularly at low $x$) and the carrier concentration that gives rise to such difference in $\rho(x,T)$. \FeCoSi\ films thus lose the low $T$ insulating behaviour of FeSi as $x$ rises.

As the temperature is reduced further below $T_\mathrm{max}$, the resistivity decreases until a minimum ($ T_\mathrm{res}$) is reached. This minimum in the resistivity curve is related to the magnetic behaviour of the films and signifies the onset of magnetic ordering in the \FeCoSi\ crystal structure.\cite{Forthaus} The position of the minimum $T_\mathrm{res}$ varies with Co doping and is found to follow the same trend as the magnetic ordering temperature $T_\mathrm{ord}$, as we shall discuss later in \S \ref{sec:mag}. Ideally, $T_\mathrm{res} \approx T_\mathrm{ord}$, but in the samples studied here, we find that $T_\mathrm{res}$ is actually slightly higher. The value of $T_\mathrm{res}$ increases with increasing Co doping and reaches the maximum value of $\sim 92$~K for $x= 0.4$ before decreasing again. The transport properties of \FeCoSi\ epilayers are dominated by short-ranged ferromagnetic interactions in the crystal structure.\citep{Onose2005} When the mean free path is of the same order as the ferromagnetic correlation length, $T_\mathrm{ord}$ and $T_\mathrm{res}$ almost coincide, as is the case for $x=0.1, 0.5$. However, if the mean free path is longer, then $T_\mathrm{res}$ is higher than $T_\mathrm{ord}$, as we observe for \FeCoSi\ films in the range $0<x<0.5$ (and discuss later in \S\ref{sec:mag}). Also this may be due to magnetic fluctuations occurring above the ordering temperature which may contribute to the discrepancy between the magnetic ordering temperature and $T_\mathrm{res}$ \citep{Pfleiderer2001} . When the temperature is decreased below $T_\mathrm{res}$, the resistivity further increases for the \FeCoSi\ films with $0 < x < 0.5$, as pointed out in the previous studies.\citep{Beille1983,Manyala2000}

Overall we observe semiconducting behaviour of the films for low $x$ and metallic for high $x$. This remains the case when the measurements were performed under a $\mu_0 H = 8$~T field applied perpendicular to the sample plane (dashed lines in Fig.~\ref{fig:TMS_RvT}). In the high temperature region (above $\sim T_\mathrm{max}$), the resistivity is almost unchanged with field for all our \FeCoSi\ films. In the lower temperature regime, after the onset of magnetic ordering, magnetoresistance gradually rises in the semiconducting regime, washing out any maximum $\rho(T, \mathrm{8~T})$. Positive magnetoresistance is a very typical property of the \FeCoSi\ system, and shall be discussed in more detail in the next section.

\section{Magnetoresistance}\label{sec:MR}

Unlike most other ferromagnetic metals, which show negative magnetoresistance (MR) at high fields,\cite{raquetprb2002} \FeCoSi\ systems show unusual positive MR in the form of bulk crystals and epilayers.\cite{Manyala2000,Onose2005,Porter2012} The high field magnetoresistance in these \FeCoSi\ samples, shown in Fig.~\ref{fig:TMS_RvT} for a perpendicular field orientation, is not only linear for $x > 0$ , but also isotropic for $ T < T_\mathrm{res}$. For an FeSi film, which is a paramagnet, the MR has a quadratic dependence on magnetic field. Introducing Co doping to FeSi, changes the nature of the curve from quadratic to linear at $x=0.1$, with a large MR ratio of almost 12\% in an 8~T field at 5~K.

\begin{figure}
  \includegraphics[width=8cm]{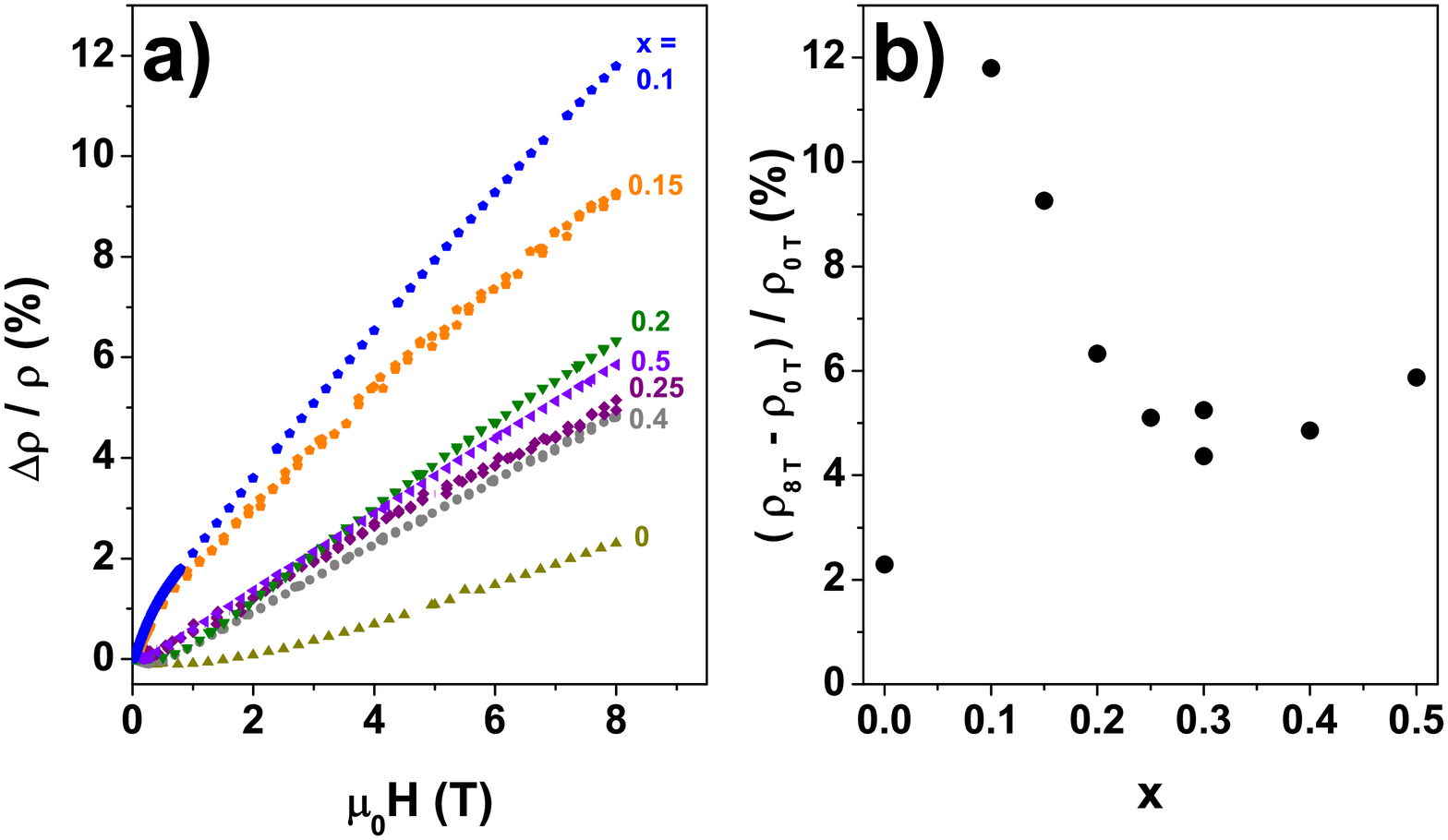}
  \caption{(Color online) Magnetoresistance. a) MR isotherms at 5~K for \FeCoSi\ films of varying Co doping $x$. b) MR ratio at 8~T and 5~K as a function of cobalt concentration $x$. \label{fig:TMS_MRvX}}
\end{figure}

Fig~\ref{fig:TMS_MRvX}(a) shows the magnetoresistance ratio observed in \FeCoSi\ epilayers for different Co doping  for a field of 8~T at 5~K. As the Co content is increased from $x=0.1$ to $x=0.5$, we observe that the MR remains linear at low temperatures ($T < T_\mathrm{res}$), i.e. in the presence of magnetic ordering. (As the temperature is increased the linearity of the MR is lost, and above $T_\mathrm{max}$ it becomes quadratic for all our \FeCoSi\ films.) The maximum magnetoresistance should be observed near the metal-insulator transition, where there is the highest Coulomb interaction. This is observed here for $x=0.1$, as shown in Fig.~\ref{fig:TMS_MRvX}(b) where we observe an MR ratio of almost 12\%. The MR ratio decreases with increasing Co content up to $x=0.3$, and then flattens off at a level of $\sim 5$\% for all higher values of $x$. The explanation of this low $T$ positive linear magnetoresistance is contested: both quantum interference effects,\cite{Manyala2000} and Zeeman splitting of the majority and minority spin bands, which reduces the high mobility minority spin carriers and in turn increases the resistivity,\cite{Onose2005} have been cited as causes.

\section{Hall effect}

\begin{figure}
  \includegraphics[width=8cm]{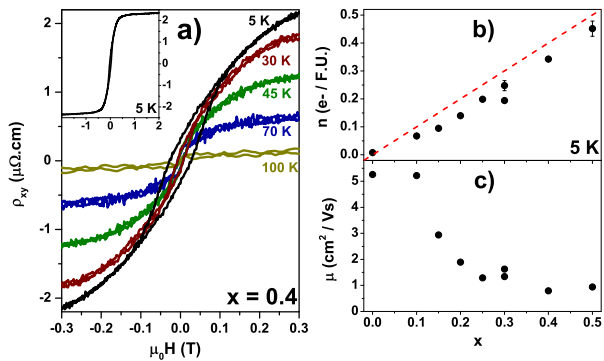}
  \caption{(Color online) Hall measurements. (a) Hall resistivity $\rho_{xy}$ as a function of field for \FeCoSi\ epilayers with $x = 0.4$ for selected temperatures. Hysteresis is observed in the extraordinary Hall effect which diminishes at elevated temperatures. The ordinary Hall effect was extracted at high fields above the saturation field. A measurement at 5~K is shown inset up to higher magnetic fields. (b) Charge carrier density expressed as electrons per formula unit inferred from measurements of the high field ordinary Hall effect at 5~K. The dashed line illustrates the ideal case of one electron added to the electron gas per cobalt atom. (c) Carrier mobility $\mu$ as a function of cobalt doping $x$ at 5~K. \label{fig:TMS_nvx}}
\end{figure}

Hall measurements were made simultaneously with the longitudinal resistivity measurements. As an example, the Hall resistivity $\rho_{xy}(H)$ for an \FeCoSi\ thin film with $x = 0.4$ is shown in Fig.~\ref{fig:TMS_nvx}(a) for various temperatures. There is low field hysteresis (for fields $\mu_0H \lesssim 0.3$~T) and a high field linear regime. (Inset in Fig.\ref{fig:TMS_nvx}(a) are data measured at 5~K showing the high field response.) The high field slope is due to the ordinary Hall effect. This high field Hall slope, measured at 5~K for \FeCoSi\ films with different values of $x$, was used to determine the type of charge carrier and carrier density, as shown in Fig. \ref{fig:TMS_nvx}(b), and was combined with the longitudinal resistivity to give the mobility of the carriers in the film, as shown in Fig.~\ref{fig:TMS_nvx}(c). In the bulk, each Co dopant contributes one conduction electron to the electron gas over the whole $x$ range.\cite{Manyala2000} The data shown in Fig.~\ref{fig:TMS_nvx}(b) show that there is a small shortfall in our samples, with close to, but not quite, one electron-like carrier per Co dopant. It is possible that there are defects in our film, too subtle to pick up by XRD or HRTEM, that act as traps preventing all the electrons released by the Co dopants from acting as carriers. As shown in Fig.~\ref{fig:TMS_nvx}(c), the mobility $\mu$ of the charge carriers drops with increasing Co doping in the films, which can be accounted for if the Co dopants act as scattering centres.

\begin{figure}
  \includegraphics[width=8cm]{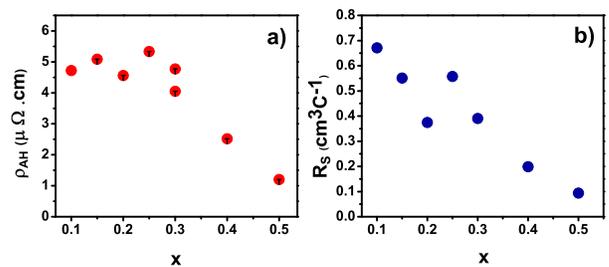}
  \caption{(Color online) Anomalous Hall effect. (a) Variation of anomalous Hall resistivity $\rho_\mathrm{AH}$, and (b) anomalous Hall coefficient $R_\mathrm{s}$ as a function of $x$ at 5~K for \FeCoSi\ films. \label{fig:TMS_AHE}}
\end{figure}

The hysteretic part of the the Hall signal arises due to the anomalous Hall effect that is present in magnetically ordered materials.\cite{Nagaosa:10} The Hall resistivity in a ferromagnetic material is given by
\begin{equation}
\rho_{xy} = R_\mathrm{o} \mu_0 H + 4\pi R_\mathrm{s} M,
\end{equation}
where $R_\mathrm{o}$ is the ordinary Hall coefficient and $R_\mathrm{s}$ is the anomalous Hall coefficient. The anomalous contribution to the Hall resistivity $\rho_\mathrm{AH} = 4\pi R_\mathrm{s} M$ was determined by extrapolating the high field Hall slope to $H = 0$, where the magnetisation is saturated, so any topological contribution of the Hall resistivity\cite{Neubauer2009,Lee2009} is neglected in the present analysis. (We will discuss it elsewhere.) $\rho_\mathrm{AH}$ for the $x=0.4$ sample, shown in Fig.~\ref{fig:TMS_nvx}(a), is as large as $2~\mu\Omega$cm at 5~K, and diminishes as $T$ rises, becoming almost negligible at 100~K or beyond. As shown in Fig.~\ref{fig:TMS_AHE}(a), even larger values of $\rho_\mathrm{AH}$ can be found for lower values of $x$. \FeCoSi\ layers with $x \lesssim 0.3$ have $\rho_\mathrm{AH} \sim 5~\mu\Omega$cm. The highest value we observe is $5.5~\mu\Omega$cm for $x=0.25$. In Fig.~\ref{fig:TMS_AHE}(b) we plot anomalous Hall coefficient $R_\mathrm{s}$ as a function of $x$ and observe that highest value is reached for $x=0.1$, up to $0.67~\pm~0.04~cm^{3} C^{-1}$ before decreasing almost linearly to $0.09~\pm~0.01~cm^{3} C^{-1}$ for $x=0.5$. The large value of $R_\mathrm{s}$ observed in our epilayers is of the similar order but a little higher than that observed in bulk \FeCoSi\ crystals by Manyala \textit{et al.}\cite{Manyala2004} This could be attributed to the strained epitaxial structure of \FeCoSi\ films, in which strain increases the effective spin-orbit coupling.

\section{Magnetic properties}\label{sec:mag}

\begin{figure}
  \includegraphics[width=8.5cm]{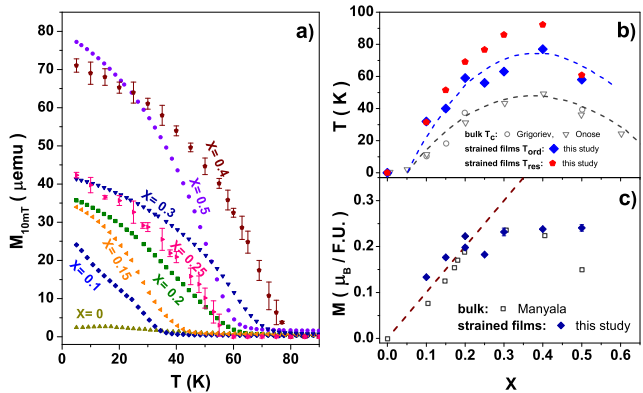}
  \caption{(Color online) Magnetic characterisation of the \FeCoSi\ epilayers.(a) Magnetisation as a function of temperature in an in-plane 10~mT field. The Co concentration, $x$, of the films is labeled on the graph. Larger error bars correspond to measurements by VSM. b) The ordering temperature $T_\mathrm{ord}$ of the epitaxial thin films shows an enhancement magnetic ordering temperature bulk material.\cite{Onose2005,Grigoriev2007} $T_\mathrm{res}$, determined as discussed in \S\ref{sec:rho}, is up to 10~K higher than $T_\mathrm{ord}$. The dashed lines are guide to the eye. (c) The saturation magnetisation at 5~K, extracted from hysteresis loops of the films, expressed in Bohr magnetons per formula unit. The value is close to $1~\mu_\mathrm{B}$ per cobalt dopant atom (ideal relationship shown by the dashed line), in good agreement with bulk,\cite{Manyala2000} for $x \lesssim 0.25$. \label{fig:TMS_Curietemp_x}}
\end{figure}

Magnetic characterisation was carried out using a vibrating sample magnetometer (VSM) with a sensitivity of $10^{-6}$ emu and a SQUID magnetometer with a sensitivity of $10^{-8}$ emu. For measurements in the VSM, several pieces of sample cut from the same wafer were stacked up to increase the signal. The temperature dependences of the magnetisation of the films were measured with a 10~mT field applied in the film plane, the results are shown in Fig.~\ref{fig:TMS_Curietemp_x}(a). It is straightforward to determine the critical temperature for magnetic ordering from these curves. Since \FeCoSi\ is helimagnetic, we refer to an ordering temperature $T_\mathrm{ord}$, rather than a Curie temperature. The values of $T_\mathrm{ord}$ obtained for the various films have been plotted as a function of Co content $x$ and shown in Fig.~\ref{fig:TMS_Curietemp_x}(b). When compared with corresponding data for bulk samples,\cite{Onose2005,Grigoriev2007} we see that for our \FeCoSi\ epilayers $T_\mathrm{ord}$ has been significantly increased, and is as high as 77~K for the $x = 0.4$ epilayer. Enhanced ordering temperatures with respect to bulk have also been observed in MnSi epilayers by Engelke \textit{et al.}\cite{Engelke2012}

We attribute this increased stability of the magnetic ordering in our \FeCoSi\ epitaxial films  to their epitaxial strain.  As shown in Fig. \ref{fig:strain_2p}(e), the biaxial in-plane strain increases the unit cell volume. Studies of bulk crystals of \FeCoSi\ under hydrostatic pressure show that compressing the unit cell volume suppresses magnetic order and can even induce a quantum phase transition in the system.\cite{Forthaus} Based on this argument, we conclude that the epitaxial strain in these \FeCoSi\ systems stabilises the magnetic order and increases $T_\mathrm{ord}$ for the whole range of $x$.

We determined the magnetic moment, in units of Bohr magnetons ($\mu_\mathrm{B}$) per formula unit(f.u.), from these hysteresis loops. The results are plotted as a function of $x$ in Fig. \ref{fig:TMS_Curietemp_x}(c). Our results are comparable to the findings of Manyala \textit{et al.} for bulk crystals,\cite{Manyala2000} and largely in line with theoretical expectations.\cite{Guevara2004} As found previously, we see that each Co atom contributes $\sim 1~\mu_\mathrm{B}$ up to a limit of $x \approx 0.25$. Beyond this point, the total moment is roughly constant at $\sim 0.25~\mu_\mathrm{B}$ per formula unit (f.u.). The dashed line in Fig. \ref{fig:TMS_Curietemp_x}(c) represents the ideal result of exactly $1~\mu_\mathrm{B}$/f.u. We can see that in the low $x$ range there is a small excess of moment per Co above the ideal result, suggesting that the Co dopants could be weakly magnetising nearby Fe atoms in this regime.

\section{Discussion and Conclusions}

In the early report of Manyala \textit{et al.}, the finding of one electron-like carrier and one $\mu_\mathrm{B}$ of magnetic moment per Co atom dopant in \FeCoSi\ (at least in the regime $x \lesssim 0.25$) was interpreted as indicating the presence of a fully spin-polarised electron gas.\cite{Manyala2000} This half-metallic state was retrodicted by band structure calculations a few years later,\cite{Guevara2004} and its presence explains the greater stability of the magnetic order against pressure for low $x$ samples.\cite{Forthaus} We previously detected evidence for the partial preservation of this state in non-phase-pure sputtered \FeCoSi\ polycrystalline films.\cite{Morley2011}

\begin{figure}[t]
  \includegraphics[width=7cm]{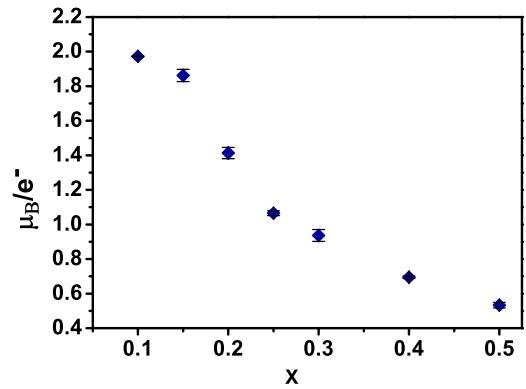}
  \caption{(Color online) Magnetic moment per carrier of the electron gas in \FeCoSi\ as a function of cobalt doping $x$.  \label{fig:spinpol}}
\end{figure}

In Fig. \ref{fig:spinpol} we show the magnetic moment per electron-like carrier as a function of $x$ for our epilayer samples. The moment is determined from the magnetometry results in Fig. \ref{fig:TMS_Curietemp_x}(c) and the number of carriers from the Hall effect, as given in Fig. \ref{fig:TMS_nvx}(b).

The data show an approximately linear decrease as the Co content $x$ rises. For $x \gtrsim 0.25$, in the metal-like regime, the behavior is much as expected: the moment per carrier ratio drops, falling to only about 0.5 $\mu_\mathrm{B}$ per electron for $x= 0.5$. The decrease in the spin-polarization for high $x$ has been previously observed and explained as being due to local disorder in the crystal structure induced by addition of Co atoms.\cite{Guevara2004,Forthaus}

In the low-doping semiconductor-like regime ($x \lesssim 0.25$), the ratio of moment per carrier exceeds unity, arising from the small shortfall in carriers per Co that was found in the data presented in Fig. \ref{fig:TMS_nvx}(b), and slight excess moment observed in Fig. \ref{fig:TMS_Curietemp_x}(c). Physically, the underlying mechanism is not clear. A plausible picture might be that there are a low number of Co atoms on Si antisites or in interstitial positions, too few to be readily detected by XRD or HRTEM, that act both as charge traps and possess local moments exceeding 1~$\mu_\mathrm{B}$ (either alone or by weakly polarising neighbouring Fe sites). More detailed studies, such as \textit{ab initio} calculations, would be required to confirm this scenario. Nevertheless, it is clear that in this regime, we have a highly spin-polarized electron gas.

To summarize, we have grown a set of \FeCoSi\ epitaxial thin films, and studied the variation in the structural, transport, and magnetic properties in the range $0 \leq x \leq 0.5$. The epilayers are $\epsilon$-phase pure, but with a deformation of the B20 unit cell into an rhombohedral form by the epitaxial strain. Qualitatively, the properties of our epilayer samples are similar in many ways to those of bulk crystals. In particular, we found the metal-insulator transition to lie in the middle of this range, with a high spin-polarization in the semiconducting regime ($x \lesssim 0.25$). However there are quantitative differences, the most important of which is the stabilisation of magnetic order up to much higher temperatures than in bulk crystals. The availability of thin films amenable to planar processing techniques is an important step to realising spintronic devices based on the remarkable physics of these B20-ordered materials.\cite{Jonietz2010,Schulz2012,Fert2013}

\begin{acknowledgments}

We acknowledge support from the EPSRC (grant EP/J007110/1) and Marie Curie Initial Training Action (ITN) Q-NET 264034. We would also like to thank Dr G. Burnell for useful discussions, Dr M. Ward for TEM sample preparation and imaging, and Dr D. Alba Venero for help with SQUID measurements.

\end{acknowledgments}

\bibliography{TMS_strain}

\end{document}